# Scattering problem for the valence electron model potential

ANZOR KHELASHVILI *, TEIMURAZ NADAREISHVILI*,**

* Academy Member, Inst. of High Energy Physics, Iv. Javakhishvili Tbilisi State University, Tbilisi, Georgia.

** Department of Physics, Faculty of Exact and Natural Sciences, , Iv. Javakhishvili Tbilisi State University,Tbilisi,

Georgia.

E-mail : anzor.khelashvili@tsu.ge ;  teimuraz.nadareishvili@tsu.ge

**ABSTRACT:**

In the paper, in the scattering problem for the valence electron model potential a self-adjoint extension is performed and Rutherford formula is modified. The scattering of slow particles for this potential is also discussed and the changes caused by the self-adjoint extension in the differential and integral cross-sections of the scattering are studied.

**Key words:** self-adjoint extension, Schrodinger equation, additional solutions, scattering amplitude, valence electron model

## I . Intoduction

The inverse squared $r^{-2}$ potential has received widespread attention in various problems of quantum mechanics [1-6]. These problems are interesting not only from an academic standpoint. A number of physically significant quantum mechanical problems manifest in such a behavior.

A detailed consideration of papers devoted to problems concerning this potential put in doubt the motivations for neglecting so-called additional (singular) solutions, which are based on mathematical sets of quantum mechanics, without invoking specific physical ideas.

The aim of this study is to consider the singular solution by using a well-known procedure of self-adjoint extension (SAE), in particular, a "pragmatic approach" [7], which is much simpler than the general Weil method but is applied only to Hamiltonians.

In [8-9] we studied the self-adjoint extension procedure for bound states in the Schrödinger equation for several analytically solvable examples (inverse-squared, valence electron and singular oscillator potentials). At the same time, it becomes necessary to carry out an analogous procedure in scattering problems, which we carried out for the inverse-squared potential in [10-11]. In this paper, a self-adjoint extension is carried out in the scattering problem for the valence electron model potential and the Rutherford formula is modified. The scattering of slow particles for this problem is also discussed. The main results of the paper are briefly summarized in the conclusion.

## II. Modification of Rutherford's formula

Consider the scattering problem in the valence electron model

$$V = -\frac{V_0}{r^2} - \frac{\alpha}{r} \quad ; \quad (V_0, \alpha > 0) \tag{2.1}$$

Note that this potential appears "naturally" in the Klein-Gordon equation for the Coulomb potential. If we follow the formalism of [12], in the scattering case we obtain the following general solution of the Schrödinger equation

$$R(r) = C_1 \rho^{-1/2+P} e^{-\frac{\rho}{2}} {}_1F_1(1/2+P-\lambda, 1+2P; \rho) + C_2 \rho^{-1/2-P} e^{-\frac{\rho}{2}} {}_1F_1(1/2-P-\lambda, 1-2P; \rho) \tag{2.2}$$

were

$$P = \sqrt{(l+1/2)^2 - 2mV_0} > 0 \tag{2.3}$$

and

$$\rho = 2ikr; \quad \lambda = -i\frac{m\alpha}{k} = -i\eta; \quad k = \sqrt{2mE}; E > 0; \quad \eta = \frac{m\alpha}{k} \tag{2.4}$$

From the behavior of the (2.2) wave function at small distances and the definition of the self-adjoint extension $\tau_S$ parameter, we obtain [10-11]

$$\tau_S = \frac{C_2(k)}{C_1(k)} (2ik)^{-2P} \tag{2.5}$$

To find the scattering amplitude, we write the representation of the (2.2) radial function at large distances. If we use the asymptotics of ${}_1F_1(kr)$ confluent hypergeometric functions [13]

$$\underset{z \to \infty}{{}_1F_1(a,c,z)} \approx e^{-i\pi a} \frac{\Gamma(c)}{\Gamma(c-a)} z^{-a} + \frac{\Gamma(c)}{\Gamma(a)} e^z z^{a-c}; \quad c \neq -n, \; n = 0,1,2... \tag{2.6}$$

and the definition of the Coulomb standard scattering amplitude [14]

$$e^{2i\delta_{Coulomb}^{st}} = \frac{\Gamma\left(\frac{1}{2}+P+\lambda\right)}{\Gamma\left(\frac{1}{2}+P-\lambda\right)} \tag{2.7}$$

We obtain

$$R \underset{r\to\infty}{\approx} e^{-\frac{\pi}{2}\eta} \frac{1}{r} e^{ikr} \left[ C_1 \frac{\Gamma(1+2P)}{\left|\Gamma\left(\frac{1}{2}+P+\lambda\right)\right|} e^{i\left(P+\frac{1}{2}\right)\frac{\pi}{2}} e^{i(X_+ + \delta^{st}_{Coulomb})} + C_2 \frac{\Gamma(1-2P)}{\left|\Gamma\left(\frac{1}{2}-P+\lambda\right)\right|} e^{-i\left(P-\frac{1}{2}\right)\frac{\pi}{2}} e^{i(X_- + \delta^{add}_{Coulomb})} \right] +$$

$$+ e^{-\frac{\pi}{2}\eta} \frac{1}{r} e^{-ikr} \left[ C_1 \frac{\Gamma(1+2P)}{\left|\Gamma\left(\frac{1}{2}+P+\lambda\right)\right|} e^{i\left(P+\frac{1}{2}\right)\frac{\pi}{2}} e^{-i(X_+ + \delta^{st}_{Coulomb})} + C_2 \frac{\Gamma(1-2P)}{\left|\Gamma\left(\frac{1}{2}-P+\lambda\right)\right|} e^{i\left(\frac{1}{2}-P\right)\frac{\pi}{2}} e^{-i(X_- + \delta^{add}_{Coulomb})} \right]$$

(2.8)

where the following notations are introduced:

$$X_+ = \eta \ln 2kr - (1/2 + P)\frac{\pi}{2}; \quad X_- = \eta \ln 2kr - (1/2 - P)\frac{\pi}{2} \qquad (2.9)$$

And, respectively, $\delta^{st}_{Coulomb}$ and $\delta^{add}_{Coulomb}$ are the Coulomb scattering amplitudes of the first (standard) and second (additional) terms of the solution to (2.1):

$$\delta^{st}_{Coulomb} = \arg \Gamma\left(\frac{1}{2}+P+\lambda\right); \quad \delta^{add}_{Coulomb} = \arg \Gamma\left(\frac{1}{2}-P+\lambda\right) \qquad (2.10)$$

On the other hand, the behavior of the $R$ radial wave function at infinity through the scattering $\delta_l$ phase shift

$$R \underset{r\to\infty}{\approx} \frac{2}{r} \sin\left(kr - \frac{l\pi}{2} + \eta \ln 2kr + \delta_l\right) \qquad (2.11)$$

(here $\eta \ln 2kr$ is the well-known term due to the slow decay of the Coulomb potential at infinity [14]), we can write as

$$R \underset{r\to\infty}{\approx} \frac{1}{ir} \left\{ e^{ikr} e^{i\left(\eta \ln 2kr - \frac{\pi}{2}l + \delta_l\right)} - e^{-ikr} e^{-i\left(\eta \ln 2kr - \frac{\pi}{2}l + \delta_l\right)} \right\} \qquad (2.12)$$

And comparing (2.8) and (2.12) gives us

$$\left[ C_1 \frac{\Gamma(1+2P)}{\left|\Gamma\left(\frac{1}{2}+P+\lambda\right)\right|} e^{i\left(P+\frac{1}{2}\right)\frac{\pi}{2}} e^{i\left(-\left(\frac{1}{2}+P\right)\frac{\pi}{2}+\delta^{st}_{Coulomb}\right)} + C_2 \frac{\Gamma(1-2P)}{\left|\Gamma\left(\frac{1}{2}-P+\lambda\right)\right|} e^{-i\left(P-\frac{1}{2}\right)\frac{\pi}{2}} e^{i\left(-\left(\frac{1}{2}+P\right)\frac{\pi}{2}+\delta^{add}_{Coulomb}\right)} \right] =$$

$$= \frac{1}{i} e^{i\left(\delta_l - \frac{\pi}{2}l + \frac{\pi}{2}\eta\right)}$$

(2.13)

$$\left[ C_1 \frac{\Gamma(1+2P)}{\left|\Gamma\left(\frac{1}{2}+P+\lambda\right)\right|} e^{i\left(P+\frac{1}{2}\right)\frac{\pi}{2}} e^{-i\left(-\left(\frac{1}{2}+P\right)\frac{\pi}{2}+\delta_{Coulomb}^{st}\right)} + C_2 \frac{\Gamma(1-2P)}{\left|\Gamma\left(\frac{1}{2}-P+\lambda\right)\right|} e^{i\left(P-\frac{1}{2}\right)\frac{\pi}{2}} e^{-i\left(-\left(\frac{1}{2}+P\right)\frac{\pi}{2}+\delta_{Coulomb}^{add}\right)} \right] =$$
$$= -\frac{1}{i} e^{-i\left(\delta_l - \frac{\pi}{2}l + \frac{\pi}{2}\eta\right)}$$
(2.14)

From which, by dividing (2.13) by (2.14) and taking into account the definition of (2.5), we obtain the $S_{VE}$ partial scattering amplitude

$$S_{VE} = e^{2i\left[l+\frac{1}{2}-P\right]\frac{\pi}{2}+2i\delta_{Coulomb}^{st}} \frac{1+\tau_S(2ik)^{2P} We^{i\left(\delta_{Coulomb}^{add}-\delta_{Coulomb}^{st}\right)}}{1+\tau_S(2ik)^{2P} We^{-2i\pi P} e^{i\left(\delta_{Coulomb}^{add}-\delta_{Coulomb}^{st}\right)}}$$
(2.15)

where the index VE indicates that the scattering occurs at the valence electron potential (2.1) and

$$W = \frac{\Gamma(1-2P)}{\Gamma(1+2P)} \frac{\left|\Gamma\left(\frac{1}{2}+P+\lambda\right)\right|}{\left|\Gamma\left(\frac{1}{2}-P+\lambda\right)\right|}$$
(2.16)

Note that in expression (2.16), the term represented by the fraction is a new term and is obtained because this time we have kept the second term in expression (2.2) and so we have performed the SAE.
(2.15) gives the correct physical results. In particular, when $\tau_S = 0$ i.e $C_2 = 0$, (2.15) gives us the standard result [17]

$$S_{VE}^{st} = e^{2i\left[l+\frac{1}{2}-P\right]\frac{\pi}{2}+2i\delta_{Coulomb}^{st}}$$
(2.17)

and for $\tau_S = \pm\infty$ i.e. $C_2 = 0$, from (2.16) we obtain an additional partial amplitude

$$S_{VE}^{st} = e^{2i\left[l+\frac{1}{2}+P\right]\frac{\pi}{2}+2i\delta_{Coulomb}^{add}}$$
(2.18)

It is necessary to note that in our work [10] we obtained the following transcendental equation for the energies of the bound states of the (2.1) potential

$$\frac{\Gamma(1/2-\lambda-P)}{\Gamma(1/2-\lambda+P)} = -\tau(-8mE)^P \frac{\Gamma(1-2P)}{\Gamma(1+2P)}$$
(2.19)

Now let us take equation (2.19) as the pole of the scattering amplitude (2.15). Here a slightly more detailed analysis is required. In particular, let us consider 3 cases.

a) $\tau_S = 0$. Standard solutions. In this case, the standard amplitude (2.17) is written using equation (2.7) as

$$S_l^{st} = e^{2i\left[l+\frac{1}{2}-P\right]\frac{\pi}{2}} \frac{\Gamma\left(\frac{1}{2}+P+\lambda\right)}{\Gamma\left(\frac{1}{2}+P-\lambda\right)}$$
(2.20)

whose poles coincide with the poles of $\Gamma(\frac{1}{2}+\lambda+P)$

$$\frac{1}{2} - \lambda + P = -n_r; \quad n_r = 0,1,2... \tag{2.21}$$

and condition (2.21), taking into account the notations (2.4), give us the standard levels obtained in the work [8]

$$E_{st} = -\frac{m\alpha^2}{2[1/2 + n_r + P]^2} = -\frac{m\alpha^2}{2\left[1/2 + n_r + \sqrt{(l+1/2)^2 - 2mV_0}\right]^2} \tag{2.22}$$

б) $\tau_S = \pm\infty$ Additional solutions. In this case, analogously to the above arguments, the poles of (2.18) will give us the $E_{add}$ additional levels obtained in [8] work.

$$E_{add} = -\frac{m\alpha^2}{2[1/2 + n_r - P]^2} = -\frac{m\alpha^2}{2\left[1/2 + n_r - \sqrt{(l+1/2)^2 - 2mV_0}\right]^2} \tag{2.23}$$

3) $\tau \neq 0, \pm\infty$. In this case (2.15) has a pole at the point

$$\tau_S (2ik)^{2P} W e^{-2i\pi P} e^{i\left(\delta_{Coulomb}^{add} - \delta_{Coulomb}^{st}\right)} = -1 \tag{2.24}$$

If we follow the usual procedure, i.e., we take the link to move to the states $k = i\mu$ i.e.

$$k^2 = -\mu^2 = 2mE; \quad (E < 0) \tag{2.25}$$

Then, taking into account (2.4), (2.16) and the following equations

$$\Gamma\left(\frac{1}{2} + P + \lambda\right) = \left|\Gamma\left(\frac{1}{2} + P + \lambda\right)\right| e^{i\delta_{Coulomb}^{st}}; \quad \Gamma\left(\frac{1}{2} - P + \lambda\right) = \left|\Gamma\left(\frac{1}{2} - P + \lambda\right)\right| e^{i\delta_{Coulomb}^{add}} \tag{2.26}$$

We obtain the formula (2.19) for the energies of the bound states.
To write the scattering phase, we use the formula [14]

$$e^{2i\tan^{-1} z} = \frac{1 + iz}{1 - iz} \tag{2.27}$$

And then (2.15) can be written as

$$S_{VE} = e^{2i\delta_{VE}} \tag{2.28}$$

Where

$$\delta_{VE} = \delta_l^{st} + \delta_{Coulomb}^{st} + arctgZ \tag{2.29}$$

The third term in the expression (2.29)

$$\delta_{SAE} = arctgZ; \quad Z = \frac{\tau_S (2k)^{2P} W \sin\left(\pi P + \delta_{Coulomb}^{add} - \delta_{Coulomb}^{st}\right)}{1 + \tau_S (2k)^{2P} W \cos\left(\pi P + \delta_{Coulomb}^{add} - \delta_{Coulomb}^{st}\right)} \tag{2.30}$$

is a new term caused by the expansion of the self-consistent expansion, and is given by the formula (2.16).
Let us now see what changes the self-extension procedure will make to Rutherford's well-known formula. As shown in [11] the elastic scattering amplitude may be rewritten as

$$f(\theta) = \frac{1}{2ik}\left\{\sum_{l=0}^{l_0-1}(2l+1)\mathcal{P}_l(\cos\theta)\left[S_{l,fall} - 1\right] + (2l_0+1)\mathcal{P}_{l_0}(\cos\theta)\right\}\left[e^{2i\left(\delta_{VE} + \tan^{-1} X_{l_0}\right)} - 1\right] +$$

$$+ \frac{1}{2ik}\sum_{l=l_0+1}^{\infty}(2l+1)\mathcal{P}_l(\cos\theta)\left[e^{2i\delta_l^{st}} - 1\right] \equiv f_1^{st}(\theta) + f_{l_0}^{SAE}(\theta) + f_2^{st}(\theta) \tag{2.31}$$

In this case, for the "fall" on the center, in the first term of (2.31), we can calculate it by our method, as was shown in [11] for the potential

$$V = -\frac{V_0}{r^2}; \quad V_0 > 0 \tag{2.32}$$

However, below we do not need this term, because we consider $l = 0$ case. In the second term of (2.31), the power of the exponent is given by (2.29).

Consider the case of small $V_0$-s, where we need to keep only the $l = 0$ member and by using following identity.

$$e^{2i(x+y)} - 1 = e^{2ix} - 1 + e^{2ix}\left(e^{2iy} - 1\right) \tag{2.33}$$

the scattering amplitude (2.31) is written as

$$f(\theta) = \frac{1}{2ik}\left\{e^{2i\left\{\left[\frac{1}{2}-P\right]\frac{\pi}{2}+\delta^{st}_{0,Coulomb}\right\}}\left[e^{2iarctgZ} - 1\right] + \sum_{l=0}^{\infty}(2l+1)P_l(\cos\theta)\left[e^{2i\left\{\left[l+\frac{1}{2}-P\right]\frac{\pi}{2}+\delta^{st}_{l,Coulomb}\right\}} - 1\right]\right\} \tag{2.34}$$

The second term of (2.34) is the standard scattering amplitude of the potential (2.1) and it is calculated in the case of small $V_0$-s in the monograph [15], but there the scattering on the Coulomb potential is considered for the Klein-Gordon equation, but as we mentioned at the beginning of this chapter, the potential (2.1) appears "naturally" in the Klein-Gordon equation for the Coulomb potential. Therefore, the results of the monograph [15] coincide with ours with the accuracy of the notations in (2.4) and so the second term of (2.34) will be

$$f^0_{VE}(\theta) = f_{Coulomb}(\theta)\left\{1 - \pi V_0 k \sin\frac{\theta}{2}e^{2i[\sigma_{-1/2}-\sigma_0]}\right\} \tag{2.35}$$

where the $f_{Coulomb}(\theta)$ Rutherford scattering amplitude is [14]

$$f_{Coulomb}(\theta) = -\frac{\eta}{2k\sin^2\frac{\theta}{2}}e^{\left\{-i\eta\ln\left(\sin^2\frac{\theta}{2}\right)+2i\sigma_0\right\}} \tag{2.36}$$

and

$$e^{2i\sigma_0} = \frac{\Gamma(1+i\eta)}{\Gamma(1-i\eta)}; \quad e^{2i\sigma_{-1/2}} = \frac{\Gamma(1/2+i\eta)}{\Gamma(1/2-i\eta)} \tag{2.37}$$

Therefore, finally (2.34) the scattering amplitude for the potential (2.1) will be

$$f(\theta) = -\frac{\eta}{2k\sin^2\frac{\theta}{2}}e^{\left\{-i\eta\ln\left(\sin^2\frac{\theta}{2}\right)+2i\sigma_0\right\}}\left\{1 - \pi V_0 k \sin\frac{\theta}{2}e^{2i[\sigma_{-1/2}-\sigma_0]}\right\} + \\ +\frac{1}{k}e^{2i\left\{\left[\frac{1}{2}-P\right]\frac{\pi}{2}+\delta^{st}_{0,Coulomb}\right\}}e^{iarctgZ}\sin arctgZ \tag{2.38}$$

Here the second term, due to the self-coupled broadening, does not depend on the $\theta$ scattering angle and for large scattering angles it becomes more pronounced when the Rutherford term itself is small; but if we take into account the terms $l \neq 0$ as well, it already becomes $\theta$ angle-dependent.

The differential scattering cross section will be

$$\frac{d\sigma}{d\Omega} = \frac{\eta^2}{4k^2 \sin^2 \frac{\theta}{2}} \left\{ 1 + (\pi V_0 k)^2 \sin^2 \frac{\theta}{2} - 2\pi V_0 k \sin \frac{\theta}{2} \cos 2\left(\sigma_{-\frac{1}{2}} - \sigma_0\right) \right\} + \frac{1}{k^2} \sin^2 Z -$$

$$-\frac{\eta \sin arctgZ}{k^2 \sin^2 \frac{\theta}{2}} \left\{ \begin{array}{l} \cos\left[\left(\frac{1}{2} - P\right)\pi + 2\delta^{st}_{0,Coulomb} + \eta \ln \sin^2 \frac{\theta}{2} - 2\sigma_0 + arctgZ\right] - \\ -\pi V_0 k \sin \frac{\theta}{2} \cos\left[\left(\frac{1}{2} - P\right)\pi + 2\delta^{st}_{0,Coulomb} + \eta \ln \sin^2 \frac{\theta}{2} - 2\sigma_{-\frac{1}{2}} + arctgZ\right] \end{array} \right\}$$

(2.39)

The last two terms in (2.39) were formed due to the self-adjoint extension procedure. The last term is especially interesting, since it depends on the sign of the parameter Z SAE defined by the formula (2.30) and therefore increases or decreases the $\frac{d\sigma_{st}}{d\Omega}$ standard cross section (when $Z = 0$) of the potential (2.1) and which may be observed in the experiment. This effect can be observed during the scattering of slow particles on the potential of the type (2.1), since at this time it is most important $l = 0$ and the situation formally resembles the inclusion of short-acting forces in the discussion.

## III. Scattering of slow particles.

Of particular interest is the scattering of slow particles, because as is well known [14] in this case the main contribution comes from the $l = 0$ states, and on the other hand, as we have noted in [8], for a potential of type

$$\lim_{r \to 0} r^2 V(r) \to -V_0 \qquad (V_0 = const > 0) \tag{3.1}$$

(to which the potential (2.1) belongs) the condition for the emergence of additional levels looks like this

$$l(l+1) < 2mV_0 < l(l+1) + 1/4 \tag{3.2}$$

i.e. for the potential (2.1) for $l = 0$ it is always necessary the SAE procedure. Let us see what changes this procedure will cause for slow particles.
For the potential (2.1) in the region

$$0 < 2mV_0 \leq 1/4 \tag{3.3}$$

we will not have a "fall" on the center (the non-falling $\tau_0$ SAE parameter is introduced according to (2.5)) and in (2.35) we keep only the first term

$$f(\theta) = \frac{1}{2ik} \left[ e^{2i\left(\delta_{0,st} + \delta^{ST}_{0,cOULOMB} + arctgz_0\right)} - 1 \right] \tag{3.4}$$

where we use formulas (2.10) and (2.30), in which we take $P_0 = \sqrt{\frac{1}{4} - 2mV_0}$ , since $l = 0$ and using formulas (2.27) we obtain

$$f(\theta) = \frac{1}{2ik} \left[ e^{2i\left(\delta_{0,st} + \delta^{ST}_{0,cOULOMB}\right)} \frac{1 + iZ_0}{1 - iZ_0} - 1 \right] \tag{3.5}$$

whence

$$\frac{d\sigma}{d\Omega} = |f(\theta)|^2 \qquad (3.6)$$

using the formula [14] for the differential section, we obtain

$$\frac{d\sigma}{d\Omega} = \frac{1}{k^2}\left\{\sin^2(\delta_{0,st} + \delta_{0,COULOMB}^{St}) + \frac{Z_0 \sin 2(\delta_{0,st} + \delta_{0,COULOMB}^{ST})}{1+Z_0^2}\right\} \qquad (3.7)$$

and for the total cross-section, which in this case coincides with the partial $\sigma_0$ one, and using the well-known formula [14]

$$\sigma_l = 4\pi(2l+1)|f_l|^2 \qquad (3.8)$$

we obtain for $l = 0$

$$\sigma = \sigma_0 = \frac{4\pi}{k^2}\left\{\sin^2(\delta_{0,st} + \delta_{0,COULOMB}^{St}) + \frac{Z_0 \sin 2(\delta_{0,st} + \delta_{0,COULOMB}^{ST})}{1+Z_0^2}\right\} \qquad (3.9)$$

Let us analyze formulas (3.7) and (3.9)

1) When $\tau_S = 0$ (ordinary quantum mechanics), then in (2.30) $Z_0 = 0$ and fom (3.7) we get

$$\frac{d\sigma}{d\Omega} = \frac{1}{k^2}\sin^2(\delta_{0,st} + \delta_{0,COULOMB}^{St}) \qquad (3.10)$$

In this formula $\delta_{0,COULOMB}^{St}$ depends on the energy as we see from (2.10), i.e. the well-known differential cross section property of the potential (2.32)

$$\frac{d\sigma}{d\Omega} = |f(\theta)|^2 \propto \frac{1}{k^2}\bigg|_{\theta=const} = \frac{1}{E}\bigg|_{\theta=const} \qquad (3.11)$$

is no longer valid [14] since this property is violated by the Coulomb potential

2) For $\tau_S = \pm\infty$, again from (2.30) we get

$$Z_0 = tg(\pi P_0 + \delta_{Coulomb}^{add} - \delta_{Coulomb}^{st}) \qquad (3.12)$$

and (3.9) gives

$$\frac{d\sigma}{d\Omega} = \frac{1}{k^2}\left\{\begin{array}{l}\sin^2(\delta_{0,st} + \delta_{0,COULOMB}^{St}) + \\ +\frac{1}{2}\sin 2(\pi P_0 + \delta_{Coulomb}^{add} - \delta_{Coulomb}^{st})\sin(\delta_{0,st} + \delta_{0,COULOMB}^{St})\end{array}\right\} \qquad (3.13)$$

From which it can be seen that in this case the formula (3.11) is still not valid.

3) (3.7) does not depend on the angle, since the second term, which is the term caused by the self-adjoint extension, does not depend on the $\theta$ scattering angle, i.e. the isotropic property characteristic of slow particles is preserved, although the potential (2.1) does not fall fast enough to infinity (it should fall faster than $r^{-3}$ [14]). If we take into account the $l \neq 0$ terms as well, it will already depend on the $\theta$ angle, i.e. the isotopic property will not be preserved.

4) From the formula (3.9) it can be seen that the total cross section depends on energy for two reasons: first, there is a multiplier $\frac{1}{k^2}$ in front and secondly, it depends on energy-dependent $Z_0$, $\delta_{Coulomb}^{st}$ quantities. Here it is already evident that the potential (2.1) does not fall fast enough to infinity and therefore the energy independence of the total cross section characteristic of slow particles is not preserved.

5) We consider the scattering of slow particles, i.e. $k \ll 1$, so we can write (2.30) as (we assume that $\tau_{S,0} k^{2P_0} W \ll 1$)

$$Z_0 = \tau_{S,0} k^{2P_0} W \sin(\pi P_0 + \delta_{Coulomb}^{add} - \delta_{Coulomb}^{st}) \qquad (3.14)$$

And then (9.7) and (9.9) give us taking into account that $Z_0^2 \ll 1$

$$\frac{d\sigma}{d\Omega} = \frac{1}{k^2} \left\{ \begin{array}{l} \sin^2(\delta_{0,st} + \delta_{0,COULOMB}^{St}) + \\ \tau_{S,0} k^{2P_0} W \sin(\pi P_0 + \delta_{Coulomb}^{add} - \delta_{Coulomb}^{st}) \sin 2(\delta_{0,st} + \delta_{0,COULOMB}^{St}) \end{array} \right\} \qquad (3.15)$$

$$\sigma = \frac{4\pi}{k^2} \left\{ \begin{array}{l} \sin^2(\delta_{0,st} + \delta_{0,COULOMB}^{St}) + \\ \tau_{S,0} k^{2P_0} W \sin(\pi P_0 + \delta_{Coulomb}^{add} - \delta_{Coulomb}^{st}) \sin 2(\delta_{0,st} + \delta_{0,COULOMB}^{St}) \end{array} \right\} \qquad (3.16)$$

From which the dependence of these quantities on the energy and the extension $\tau_{S,0}$ parameter is clearly visible. It is clear that the $\sigma > 0$ physical requirement will impose the following restriction on $\tau_{S,0}$

$$\tau_{S,0} > -\frac{tg(\delta_{0,st} + \delta_{0,COULOMB}^{St})}{k^{2P_0} W \sin(\pi P_0 + \delta_{Coulomb}^{add} - \delta_{Coulomb}^{st})} \qquad (3.17)$$

## IV. CONCLUSIONS

In the paper, in the scattering problem for the valence electron model potential a self-adjoint extension is performed and Rutherford formula is modified formula. In particular, it is shown that in the differential scattering cross-section formula, due to the SAE procedure, two new terms arise, which depend on the sign of the self-adjoint extension parameter and by this reason increases or decreases the $\frac{d\sigma_{st}}{d\Omega}$ standard cross section of the potential (2.1) and which may be observed in the experiment.

The scattering of slow particles for this potential is also considered, and the changes caused by the self-adjoint extension in the representations of the differential and integral scattering cross-sections are studied.